\documentclass[aps,reprint,superscriptaddress,twocolumn,prl,nobalancelastpage]{revtex4-2}
\usepackage{amsmath}
\usepackage{amsfonts}
\usepackage{amssymb}
\usepackage{graphicx}
\usepackage{xcolor}
\usepackage{sidecap}
\usepackage[colorlinks=True,citecolor=myRed,linkcolor=myRed,urlcolor=myBlue]{hyperref}
\usepackage{hypcap}
\usepackage{braket}
\usepackage{yfonts}
\usepackage{bm}
\usepackage[mathscr]{eucal}
\usepackage{marginnote}
\usepackage[normalem]{ulem}
\newcommand\eq[1]{\begin{align}#1\end{align}}
\newcommand\tcrit{\theta_{c,\mathrm{crit}}}
\definecolor{myBlue}{RGB}{113,73,75}
\definecolor{myOrange}{RGB}{229,131,46}
\definecolor{myGreen}{RGB}{44,160,44}
\definecolor{myRed}{RGB}{155,57,58}
\definecolor{myPurple}{RGB}{148,103,189}

\makeatletter
\def\p@figure{\color{myBlue}}
\def\p@equation{\color{myRed}}
\makeatother

\begin{document}
\title{Constrained Dynamics and Directed Percolation}

\author{Aydin Deger}
\email{a.deger@lboro.ac.uk}
\affiliation{Interdisciplinary Centre for Mathematical Modelling and Department of Mathematical Sciences,
  Loughborough University, Loughborough, Leicestershire LE11 3TU,
  UK}

\author{Achilleas Lazarides}
\email{a.lazarides@lboro.ac.uk}
\affiliation{Interdisciplinary Centre for Mathematical Modelling and Department of Mathematical Sciences,
  Loughborough University, Loughborough, Leicestershire LE11 3TU,
  UK}

\author{Sthitadhi Roy}
\email{sthitadhi.roy@icts.res.in}
\affiliation{International Centre for Theoretical Sciences, Tata Institute of Fundamental Research, Bengaluru 560089, India}
\date{\today}

\begin{abstract}
    In a recent work [\href{https://journals.aps.org/prl/abstract/10.1103/PhysRevLett.129.160601}{A. Deger \textit{et al.}, Phys. Rev. Lett. \textbf{129}, 160601 (2022)}] we have shown that kinetic constraints can completely arrest many-body chaos in the dynamics of a classical, deterministic, translationally-invariant spin system with the strength of the constraint driving a dynamical phase transition. Using extensive numerical simulations and scaling analyses we demonstrate here that this constraint-induced phase transition lies in the directed percolation universality class in both one and two spatial dimensions.
\end{abstract}

\maketitle

Kinetic constraints have long emerged as a prominent avenue towards impeding ergodicity~\cite{fredrickson1984kinetic,fredrickson1985facilitated,jackle1991hierarchically,sollich1999glassy,sollich2003glassy,ritort2003glassy,garrahan2011kinetically}, complementary to and fundamentally different from that of quenched randomness and originally introduced in the context of the dynamics of glass forming systems at low temperatures. The underlying mechanism of dynamical  slowing down is the decrease of the effective connectivity in configuration space by forbidding processes based on local constraints~\cite{roy2020strong}. This is to be contrasted to the mechanisms of glassy relaxation in disordered systems, involving topographic features of the potential energy landscape. In many-body quantum systems it has been established in the last few years that constraints can lead to slow relaxation and also stabilise a many-body localised phase, not only at low temperatures but also at infinite temperatures~\cite{horssen2015dynamics,lan2018quantum,pancotti2020quantum,roy2020strong}.

This motivates the following question: what is the fate of \textit{classical} many-body chaos in the presence of constrained dynamics? {A signature of chaos is that infinitesimally small perturbations grow exponentially, dramatically changing the state at later times. In extended many-body systems and for spatially localised perturbations, this effect spreads ballistically in space, eventually resulting in global changes in the state~\cite{das2018light}. We have recently showed that constraints in the dynamics can, if strong enough, fully arrest this spreading, confining the effect of a perturbation to a small region and thus suppressing chaos~\cite{deger2022arresting}.} The nature of the transition between these chaotic and frozen phases, in particular its universality class, has however remained an open question. 

% This motivates the following question: what is the fate of \textit{classical} many-body chaos in the presence of constrained dynamics? \textcolor{blue}{A telltale signature of chaos is that infinitesimal perturbations in initial conditions amplify exponentially such that the trajectory of the system at later times is wildly different from that with the unperturbed initial conditions. Additionally, in extended many-body systems,  an infinitesimal \emph{local} perturbation spreads ballistically in space to create global changes in the state at later times~\cite{das2018light}. Recently however, we showed that constraints in the dynamics, if strong enough, can fully arrest the spreading of the perturbation~\cite{deger2022arresting}. In fact, the strength of the constraint can drive a dynamical phase transition separating a chaotic phase from a localised one, where the effect of the perturbation remains spatially confined~\cite{deger2022arresting}.} 
% %Specifically, we track the spreading of a perturbation across the system, finding ballistic spreading in the chaotic phase whereas in the arrested phase the effect of the perturbation remains spatially confined \cite{deger2022arresting}. 
% The precise nature of the transition, in particular with regard to its universality class however, remained an open question.

Via numerical simulations and scaling analyses, we provide a sharp answer to this question: the phase transition belongs to the directed percolation (DP) universality class~\cite{kinzel1983percolating,kinzel1985phase,hinrichsen2000non,henkel2008non}. This constitutes the central result of this work, and is remarkable because the models we consider are translation-invariant, with deterministic dynamical rules, quite differently from conventional percolation models. To the best of our knowledge, the appearance of DP universality in such a clean, deterministic setting has not been reported before.

{The DP problem is an anisotropic variant of the standard percolation problem such that the percolating cluster can grow only in a given direction~\cite{broadbent1957percolation}. Equivalently, it can be understood as a cluster growth process on a graph with bonds directed along the given direction. This naturally endows the problem with a dynamical interpretation with time as one of the directions of the graph along which the bonds are directed.}
In order to map our problem onto a DP problem, we note that at late times, in the non-chaotic phase the spins freeze while in the chaotic phase they remain dynamically active. Based on this observation, we map the dynamically active and frozen spins to occupied and empty sites on the space-time lattice for the DP problem.

\begin{figure}
\includegraphics[width=\linewidth]{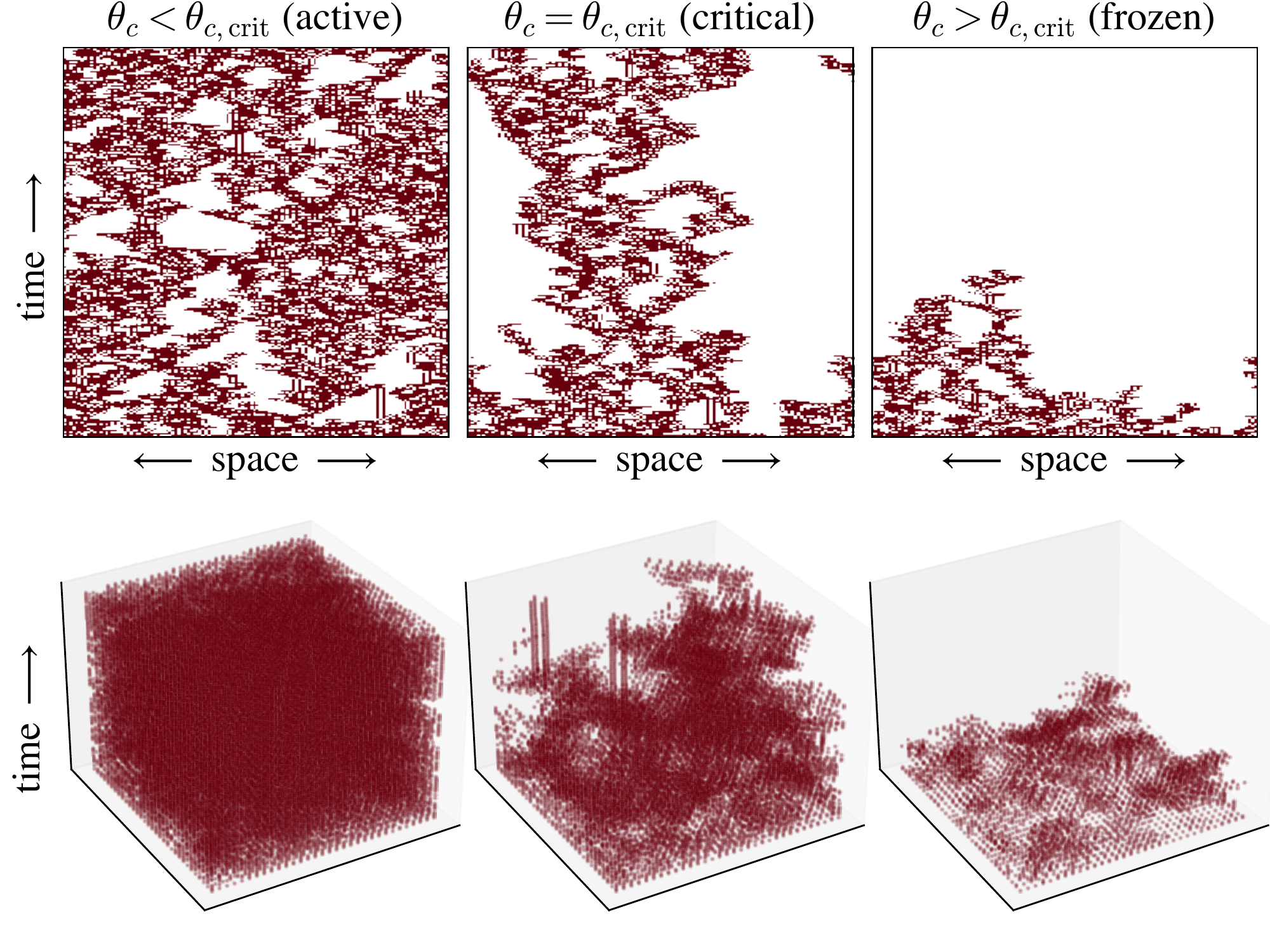}
\caption{Instances of active clusters for constrained spin dynamics in the active phase (left), at criticality (middle), and in the frozen phase (right) in the space-time graph in 1+1D (top) and 2+1D (bottom). In all cases we start from an all-sites-active configuration. In the active phase, there is always a finite density of active sites at arbitrarily long times whereas in the frozen phase the system goes into an absorbing state where all spins are frozen.}
\label{fig:clusters}
\end{figure}

We consider constrained dynamics of a periodically driven classical Heisenberg spin system in spatial dimensions $d=1$ and $d=2$.  Driving ensures that this model has no conserved quantities, including total energy. The Hamiltonian within a period, $T$, is given by
\eq{
    H(t)=\begin{cases}
    \sum_{\braket{i,j}}S_i^zS_j^z + h\sum_i S_i^z;~&t\in[0,T/2)\\
    g\sum_i S_i^x;~&t\in[T/2,T)
    \end{cases}\,,
    \label{eq:ham}
}
where $\braket{i,j}$ denotes a pair of nearest-neighbour sites. The equations for the stroboscopic dynamics of the spins in the presence of kinetic constraints are then
\eq{
    \vec{S}_i(t+T) = \mathsf{R_x}[\gamma_{x,i}(t)]\cdot \mathsf{R_z}[\gamma_{z,i}(t)]\cdot \vec{S}_i(t)\,,
    \label{eq:rotation}
}
where $\mathsf{R_{x(z)}}[\gamma_{x(z),i}]$ denotes rotation matrices about the $x(z)$ axis by an angle $\gamma_{x(z),i}$. These angles are given by
\eq{
    \begin{split}
    \gamma_{z,i}(t) &= \Theta_i(t)\bigg[\sum_{j\in\braket{i,j}}S^z_j(t)+h\bigg]T/2\\
    \gamma_{x,i}(t) &= \Theta_i(t)g T/2\\
    \end{split}\,
    \label{eq:angles}
}
where $\Theta_i(t)$  encodes the kinetic constraint via a Heaviside step function
\eq{
    \Theta_i(t) = \Theta[\cos\theta_c-\min_{j|j\in\braket{i,j}}S^z_j(t)]\,.
    \label{eq:constraint}
}
This constraint means that the spin at site $i$ rotates under the dynamics only if at least one of its neighbouring spins lies outside the spherical sector subtended by a polar angle $\theta_c$. We will call such a spin \textit{active}. Corollarily, a spin is \textit{frozen} and does not evolve dynamically if all its neighbours lie inside the spherical sector. {The constraint \eqref{eq:constraint} is inspired by the Fredrickson-Andersen model of constrained Ising spin glasses~\cite{fredrickson1984kinetic,fredrickson1985facilitated}. The physics is that dynamics is locally forbidden in a region if it is surrounded by immobile high-density regions, modelled by up spins and allowed if there are some mobile low-density regions, modelled by down spins, in the neighbourhood. We generalise this via Eq.~\eqref{eq:constraint} to the case of Heisenberg spins.}
The angle $\theta_c$ therefore parametrises the strength of the constraint and, as we will show, tunes the system across a dynamical phase transition at $\tcrit$ between an \textit{active} phase at $\theta_c<\tcrit$ and a \textit{frozen} phase at $\theta_c>\tcrit$. To distinguish between these and to map the problem to DP, which is usually discussed in terms of binary variables, we define an indicator function $\sigma_i(t)$, which we call the \emph{activity}. It takes a value 1 if the spin at site $i$ is active at time $t$ and $0$ otherwise; in other words, $\sigma_i(t)=\Theta_i(t)$. In terms of this, we define the density of active sites at time $t$ as
\eq{
    \rho(t) = N^{-1}\sum\nolimits_{i}\sigma_i(t)\,,
    \label{eq:rhot}
}
where $N$ is the total number of spins.

Numerically simulating the dynamics starting from an all-active initial state, we find that in the active phase, $\theta_c<\tcrit$, there is always a finite density of active sites at arbitrarily long times, as illustrated in Fig.~\ref{fig:clusters}(left). This implies that the long-time state is evolving dynamically and fluctuating. On the other hand, in the frozen phase, $\theta_c>\tcrit$, the system goes into a state at late times wherein the $\sigma_i$ stops fluctuating with $t$. In order words, the state described in terms of $\sigma_i$ gets \textit{absorbed} into a frozen one. The constraint-induced dynamical phase transition is therefore an \textit{absorbing phase transition}. We find that, in the frozen phase, but near the critical point, the typical absorbing state is one where $\sigma_i=0$ for all sites [Fig.~\ref{fig:clusters}(right)]. Once the system reaches such a state, since $\Theta_i=0$ for all $i$ in Eqs.~\ref{eq:rotation} and \ref{eq:angles}, the dynamics is completely frozen. 

We also find other absorbing states where there exist temporally persistent, spatially finite and dynamically stable configurations surrounded by inactive spins--reminiscent of breathers. These evolve regularly without spreading and persist forever. However, due to the extreme diluteness of such active sites, they are statistically irrelevant for the scaling behaviour near the critical point~\cite{supp}. We can therefore posit that $\rho(t)$ in the limit of $t\to\infty$ is a valid order parameter, with the active and frozen phases characterised by $\rho_\infty\equiv \rho(t\to\infty)\to$ finite and vanishing values respectively.

\begin{figure*}
\includegraphics[width=\linewidth]{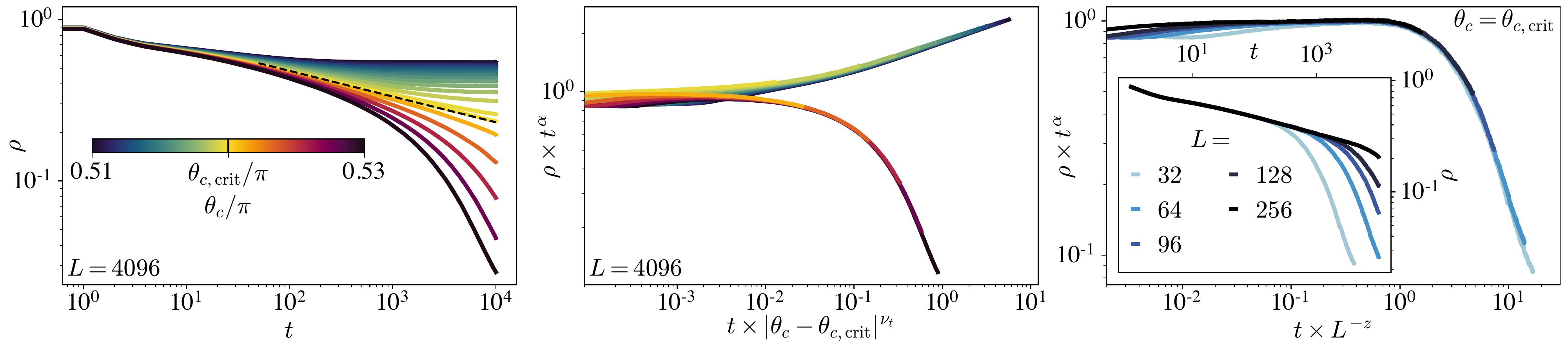}
\caption{Critical scaling in the 1+1D case. Left: The density of active sites, $\rho$, as a function of $t$ for different values of $\theta_c$. At the critical point, $\rho$ decays with $t$ following a power law with an exponent $\alpha=0.159$ consistent with DP in 1+1D (black dashed line). Middle: Scaling $\rho$ in the left panel with $t^\alpha$ and plotting it against $t|\theta_c-\tcrit|^{\nu_t}$ with $\theta_{c,\mathrm{crit}}\approx 0.5234\pi$ shows perfect scaling collapse for the DP universality exponents of $\alpha=0.159$ and $\nu_t=1.734$. Right: Finite-size scaling at the critical point by plotting $\rho t^\alpha$ against $tL^{-z}$ with te DP universality exponent, $z=1.581$, again shows excellent collapse. The inset shows the unscaled data. Results for $g=0.4$, $h=0.1$, and $T=2\pi$.}
\label{fig:1dheis}
\end{figure*}

Finally, for the dynamics of $\sigma_i(t)$ to be a bonafide DP problem, we need to argue that the active sites necessarily form a contiguous cluster in the space-time graph connecting all active sites at time $t$ to the initially-active sites at $t=0$. In other words, that the dynamics cannot spawn active clusters in a background of frozen sites. This is straightforwardly argued for based on the locality of the constraints: a spin can change state over a period if and only if at least one of its neighbours is active. Therefore if a spin is inactive, $\sigma_i(t)=0$ it may only become active $\sigma_i(t+T)=1$ if either of $\sigma_{i\pm 1}(t)=1$. This implies that any active spin $\sigma_i(t)=1$ has one of its parents active, $\sigma_{i\pm 1}(t-T)=1$, and so on up to the initial time $t=0$.  This implies that the active sites form a contiguous cluster in the space-time graph. Therefore, the active phase of the constrained dynamics corresponds to the percolating phase as the cluster of active sites percolates all the way to infinite time whereas the frozen phase corresponds to the non-percolating phase as the cluster of active sites dies out.

From the above discussion, we conclude that the constraint-induced dynamical phase transition can be described as a continuous, absorbing phase transition with a one-component order parameter, short-ranged dynamical rules and no symmetries except translation invariance. It therefore satisfies three out of the four conjectured requirements by Janssen and Grassberger~\cite{janssen1981nonequilibrium,grassberger1982phase} for the transition to be in the DP universality class. The one requirement that our model does not satisfy is the presence of a unique absorbing state (any configuration of the spins inside the cones is an absorbing state). Nevertheless the DP universality is known to be extremely robust against such violations of the aforementioned requirements~\cite{jensen1993critical,jensen1994critical,albano1995irreversible,munoz1996critical,munoz1997infinite,munoz1998phase}. In the following, using extensive numerical simulations and scaling analyses we firmly establish that our constraint-induced transition does indeed lie in the DP universality class.

Before delving into the results, let us briefly recapitulate the scaling forms and the critical exponents for the DP universality class. Since the DP transition is a continuous phase transition, the order parameter goes to zero continuously with an exponent $\beta$ from the active side as
\eq{
\rho_\infty\sim \Delta^{\beta}\,\quad \Delta\equiv\tcrit-\theta_c\,.
\label{eq:beta}
}
In addition to a correlation length, $\xi_x$, diverging as $\xi_x\sim |\Delta|^{-\nu_x}$, we also have a correlation time, $\xi_t$, which diverges with a different exponent $\xi_t\sim|\Delta|^{-\nu_t}$. This also defines the dynamical exponent $z=\nu_t/\nu_x$ which relates the rescaling of space and time under rescaling of the parameter $\Delta$ that tunes the phase transition. The DP transition is thus described in terms of the three independent critical exponents $(\beta,\nu_x,\nu_t)$, which are strictly defined in the steady state. Since the true steady state is not accessible in numerical calculations, we obtain these exponents from dynamical scaling as follows. With an initial condition where all sites are active, scale invariance at the critical point suggests $\rho(t,\tcrit)$ decaying as a power law for an infinite system, $\rho(t,\tcrit)\sim t^{-\alpha}$. Usual considerations of critical scaling imply that corrections away from this limit are captured via universal scaling functions of $t/\xi_t\sim t|\Delta|^{\nu_t}$ and of $t/L^z$,
\eq{
    \rho(t) \sim t^{-\alpha}f(t|\Delta|^{\nu_t},tL^{-z})\,.
    \label{eq:dynamic-scaling}
}
where $L$ is the linear size of the system and $N=L^d$ with $d$ with the spatial dimension.

For very large systems such that $tL^{-z}\ll 1$, asymptotically in the active phase such that $t\gg\xi_t$, $\rho(t)$ saturates to a constant and hence we expect the scaling function $f$ in Eq.~\eqref{eq:dynamic-scaling} to be such that the time-dependence in $\rho(t)$ drops out in this limit. This implies $f(y_1,y_2\ll 1)\sim y_1^{\alpha}$ for $y_1\gg 1$ and hence $\rho(t\to\infty)\sim \Delta^{\alpha\nu_t}$ in the active phase. Comparing this to Eq.~\eqref{eq:beta}, we find the relation between the exponents $\alpha = \beta/\nu_t$.
Therefore, by performing a scaling analyses on the data for $\rho(t)$, one can extract the exponents $\alpha$, $\nu_t$, and $z$ and hence the three fundamental exponents $\beta$, $\nu_x$, and $\nu_t$. Table~\ref{tab:exponents} summarises the known values of these exponents.
\begin{table}
\begin{center}
\begin{tabular}{c||c|c|c|c|c}
    $d$/exponents & $\beta$ & $\nu_x$ & $\nu_t$ & $\alpha$ & $z$\\
    \hline\hline
    $d=1$ & 0.276 & 1.097 & 1.734 & 0.159 & 1.581 \\
    $d=2$ & 0.584 & 0.734 & 1.295 & 0.451 & 1.76
\end{tabular}
\end{center}
\caption{Summary of the DP universality class critical exponents for $d=1$ and $d=2$ taken from Ref.~\cite{hinrichsen2000non}.}
\label{tab:exponents}
\end{table}

\begin{figure}
\includegraphics[width=\linewidth]{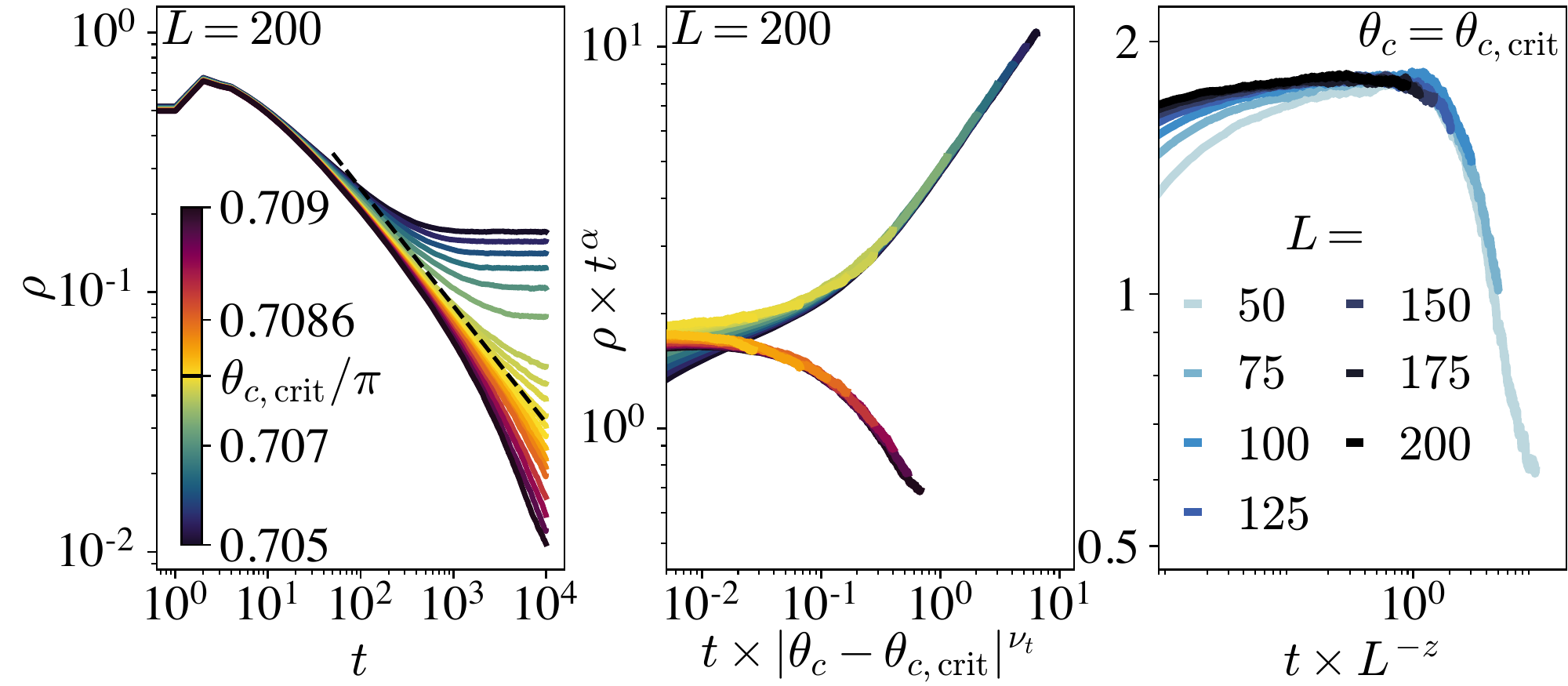}
\caption{Critical scaling in the 2+1D case. Analogous figure to Fig.~\ref{fig:1dheis} but for $d=2$. In the left panel, the dashed black line corresponds to $t^{-\alpha}$ with $\alpha=0.451$. The middle panel shows the collapse of $\rho t^{\alpha}$ onto a universal functions of $t|\theta_c-\tcrit|^{\nu_t}$ with $\theta_{c,\mathrm{crit}}\approx 0.7084\pi$ and $\nu_t=1.295$, consistent with DP universality. Right panel shows finite-size scaling collapse at $\tcrit$ of $\rho t^\alpha$ as function of $tL^{-z}$ with $z=1.76$, the DP universality value. Results for $g=0.4$, $h=0.1$, and $T=2\pi$.}
\label{fig:2dheis}
\end{figure}

Let us now turn to our results for the constrained spin dynamics described via Eq.~\eqref{eq:ham} through Eq.~\eqref{eq:constraint} for both $d=1$ and $d=2$. We evolve our system using the full dynamics for $\{\vec{S}_i(t)\}$, then calculate $\{\sigma_i(t)\}$ and then $\rho(t)$. For the former, we consider a chain and for the latter, a square lattice.  The results for the 1+1D case are shown in Fig.~\ref{fig:1dheis}
whereas those for the 2+1D case in Fig.~\ref{fig:2dheis}. Our initial conditions are chosen randomly except for ensuring that all spins are active at $t=0$. This is done by initialising randomly the polar and azimuthal angles of the spins, $\arccos(S^z_i)$ and $\arctan(S^x_i/S^y_i)$, from uniform distributions $\in(\theta_c,\pi]$ and $\in[0,2\pi)$ respectively. {In what follows, we set $T=2\pi$, $h=0.1$, and $g=0.4$ without loss of generality.}

In the left panels we show $\rho(t)$, defined in Eq.~\eqref{eq:rhot}, as a function of $t$ for different values of $\theta_c$ straddling $\tcrit$ for the largest sizes in our simulations; $L=4096$ for $d=1$ and $L=200$ for $d=2$. For $\theta_c<\tcrit$, the data on logarithmic axes curves upwards from a power-law indicating its tendency to saturate to a finite value at $t\to\infty$, signifying the active/percolating phase. On the other hand, for $\theta_c>\tcrit$, the deviation of $\rho(t)$ from the power-law curves downwards indicating a rapid decay of $\rho(t\to\infty)\to 0$, a signature of the frozen/non-percolating phase. We therefore estimate the critical point from the $\rho(t)$ data as the $\theta_c$ value which has the minimal curvature. At the so-estimated, $\theta_c=\tcrit$, the data shows a perfect power-law indicated by the straight line with almost no curvature on logarithmic axes. The black dashed lines show the expected power-laws from the DP universality exponents (see Table~\ref{tab:exponents}). They are in excellent agreement with the data which in turn confirm that the $\alpha$ exponent is the same as that of DP universality.

The sizes we considered ensure $tL^{-z}\ll 1$ and hence the $L$-dependence in the scaling function \eqref{eq:dynamic-scaling} can be ignored. We confirm this by ensuring the data in Figs.~\ref{fig:1dheis}(left) and \ref{fig:2dheis}(left) are converged with $L$. In this limit $\rho(t)t^\alpha$ is a universal function of $t|\theta_c-\tcrit|^{\nu_t}$. Upon rescaling $\rho(t)$ with $t^\alpha$ and plotting it against $t|\theta_c-\tcrit|^{\nu_t}$ with $\alpha$ and $\nu_t$ from Table~\ref{tab:exponents} and $\tcrit$ extracted as above, we find that the data for all $\theta_c$ collapse onto two universal curves, one for each phase. This is shown in the middle panels in Figs.~\ref{fig:1dheis} and \ref{fig:2dheis} for $d=1$ and $d=2$ respectively, which confirms that the $\nu_t$ exponent is also the same as that of DP universality.

Finally, we consider the finite-size scaling at the critical point to extract the dynamical exponent $z$. At criticality, $\Delta=0$, and hence the scaling function \eqref{eq:dynamic-scaling} implies that $\rho(t)t^\alpha$ is an universal function of $tL^{-z}$. In the right panels of Figs.~\ref{fig:1dheis} and \ref{fig:2dheis}, we plot $\rho(t)t^\alpha$ as a function of $tL^{-z}$ at $\theta_c=\tcrit$ with $z$ from Table~\ref{tab:exponents} and find that the curves for several $L$ collapse on top each other. This confirms that the dynamical exponent $z$  is also the same as that of DP universality.

The analyses presented in the three panels together in Fig.~\ref{fig:1dheis} for $d=1$ and Fig.~\ref{fig:2dheis} for $d=2$ thus show that the three exponents $\alpha$, $\nu_t$ and $z$, and hence by extension the three independent exponents $\beta$, $\nu_x$, and $\nu_t$ for our constraint-induced dynamical phase transition are the same as those for DP universality class. We therefore conclude that the transition lies in the same universality--this constitutes the central result of this work.

Note that the dynamics in our model are completely translation-invariant and deterministic. In the past absorbing transitions in deterministic systems were found to have non-universal model-dependent critical exponents~\cite{houlrik1990meanfield}, which makes the appearance of DP universality in our case remarkable. To intuitively understand this result, we introduce a stochastic cellular automaton (CA). This is to be viewed as a coarse-grained version of the spin system, having broadly the same features and constraints in its dynamics. This CA involves a Boolean variable per site, $\tau_i$, representing whether the spin at $i$ is inside (0)/outside (1) the cone. The dynamical rule is then that $\tau_j(t+1)=\tau_j(t)$ if both parents $\tau_{j\pm 1}(t)=0$ or $\tau_j(t+1)$ is 1 or 0 with probabilities $1-p$ and $p$ if either or both $\tau_{j\pm1}(t)=1$. Numerically, we find this CA to display a transition in the DP class~\cite{supp}. Note that this occurs even though our simplified CA model replaces the continuous spin degree of freedom with a two-state variable and the deterministic by stochastic dynamics, and in so doing completely misses spatiotemporal correlations in (or, history dependence of) the dynamics.

Our work should also be connected to and contrasted with transitions belonging to the DP universality found in certain cellular automatons in a different setting~\cite{liu2021butterfly,willsher2022measurement}. Apart from the fundamental difference from our work that the dynamics in these models was stochastic, the mapping to DP was based on two copies of the system and the active/inactive sites were defined based on whether the Ising variables were different/same in the two copies. In this sense it was a damage spreading process~\cite{kauffman1984emergent,martin1985lyapunov,derrida1986phase}, while in our work the mapping requires only one copy of the system and the growing DP cluster corresponds to a growing cluster of dynamically active.

To summarise, we have established that the constraint-induced dynamical phase transition in clean, deterministic systems from a chaotic phase to one where the dynamics is arrested~\cite{deger2022arresting} can be mapped onto a DP problem and the transition lies in  the conventional DP universality class. A question of interest for future work is under what conditions such transitions may fall out of this universality class. A natural setting to look for such cases would be systems with long-ranged power-law interactions and constraints. In particular, one may write a model where the constraints, such as those in Eq.~\eqref{eq:constraint} are long-ranged -- then an active site at $i$ can activate another at $i+r$ with a probability with a strength which falls off as a power-law with $r$. Whether such a model has a transition or not, and if it does, does it share features with the DP transition in long-ranged spreading processes~\cite{grassberger1986spreading,janssen1999levy,hinrichsen1999model} is a question of interest.

\begin{acknowledgments}
AD and AL acknowledge support from EPSRC Grant No. EP/V012177/1 and SR acknowledges support from an ICTS-Simons Early Career Faculty Fellowship by a grant from the Simons Foundation (677895, R.G.). The numerical simulations were performed on the Contra cluster at ICTS-TIFR.
\end{acknowledgments}

\bibliography{refs}

%apsrev4-2.bst 2019-01-14 (MD) hand-edited version of apsrev4-1.bst
%Control: key (0)
%Control: author (8) initials jnrlst
%Control: editor formatted (1) identically to author
%Control: production of article title (0) allowed
%Control: page (0) single
%Control: year (1) truncated
%Control: production of eprint (0) enabled
\begin{thebibliography}{36}%
\makeatletter
\providecommand \@ifxundefined [1]{%
 \@ifx{#1\undefined}
}%
\providecommand \@ifnum [1]{%
 \ifnum #1\expandafter \@firstoftwo
 \else \expandafter \@secondoftwo
 \fi
}%
\providecommand \@ifx [1]{%
 \ifx #1\expandafter \@firstoftwo
 \else \expandafter \@secondoftwo
 \fi
}%
\providecommand \natexlab [1]{#1}%
\providecommand \enquote  [1]{``#1''}%
\providecommand \bibnamefont  [1]{#1}%
\providecommand \bibfnamefont [1]{#1}%
\providecommand \citenamefont [1]{#1}%
\providecommand \href@noop [0]{\@secondoftwo}%
\providecommand \href [0]{\begingroup \@sanitize@url \@href}%
\providecommand \@href[1]{\@@startlink{#1}\@@href}%
\providecommand \@@href[1]{\endgroup#1\@@endlink}%
\providecommand \@sanitize@url [0]{\catcode `\\12\catcode `\$12\catcode
  `\&12\catcode `\#12\catcode `\^12\catcode `\_12\catcode `\%12\relax}%
\providecommand \@@startlink[1]{}%
\providecommand \@@endlink[0]{}%
\providecommand \url  [0]{\begingroup\@sanitize@url \@url }%
\providecommand \@url [1]{\endgroup\@href {#1}{\urlprefix }}%
\providecommand \urlprefix  [0]{URL }%
\providecommand \Eprint [0]{\href }%
\providecommand \doibase [0]{https://doi.org/}%
\providecommand \selectlanguage [0]{\@gobble}%
\providecommand \bibinfo  [0]{\@secondoftwo}%
\providecommand \bibfield  [0]{\@secondoftwo}%
\providecommand \translation [1]{[#1]}%
\providecommand \BibitemOpen [0]{}%
\providecommand \bibitemStop [0]{}%
\providecommand \bibitemNoStop [0]{.\EOS\space}%
\providecommand \EOS [0]{\spacefactor3000\relax}%
\providecommand \BibitemShut  [1]{\csname bibitem#1\endcsname}%
\let\auto@bib@innerbib\@empty
%</preamble>
\bibitem [{\citenamefont {Fredrickson}\ and\ \citenamefont
  {Andersen}(1984)}]{fredrickson1984kinetic}%
  \BibitemOpen
  \bibfield  {author} {\bibinfo {author} {\bibfnamefont {G.~H.}\ \bibnamefont
  {Fredrickson}}\ and\ \bibinfo {author} {\bibfnamefont {H.~C.}\ \bibnamefont
  {Andersen}},\ }\bibfield  {title} {\bibinfo {title} {Kinetic {I}sing model of
  the {G}lass {T}ransition},\ }\href
  {https://doi.org/10.1103/PhysRevLett.53.1244} {\bibfield  {journal} {\bibinfo
   {journal} {Phys. Rev. Lett.}\ }\textbf {\bibinfo {volume} {53}},\ \bibinfo
  {pages} {1244} (\bibinfo {year} {1984})}\BibitemShut {NoStop}%
\bibitem [{\citenamefont {Fredrickson}\ and\ \citenamefont
  {Andersen}(1985)}]{fredrickson1985facilitated}%
  \BibitemOpen
  \bibfield  {author} {\bibinfo {author} {\bibfnamefont {G.~H.}\ \bibnamefont
  {Fredrickson}}\ and\ \bibinfo {author} {\bibfnamefont {H.~C.}\ \bibnamefont
  {Andersen}},\ }\bibfield  {title} {\bibinfo {title} {Facilitated kinetic
  {I}sing models and the glass transition},\ }\href
  {https://aip.scitation.org/doi/10.1063/1.449662} {\bibfield  {journal}
  {\bibinfo  {journal} {J. Chem. Phys.}\ }\textbf {\bibinfo {volume} {83}},\
  \bibinfo {pages} {5822} (\bibinfo {year} {1985})}\BibitemShut {NoStop}%
\bibitem [{\citenamefont {J{\"a}ckle}\ and\ \citenamefont
  {Eisinger}(1991)}]{jackle1991hierarchically}%
  \BibitemOpen
  \bibfield  {author} {\bibinfo {author} {\bibfnamefont {J.}~\bibnamefont
  {J{\"a}ckle}}\ and\ \bibinfo {author} {\bibfnamefont {S.}~\bibnamefont
  {Eisinger}},\ }\bibfield  {title} {\bibinfo {title} {A hierarchically
  constrained kinetic {I}sing model},\ }\href
  {https://doi.org/https://doi.org/10.1007/BF01453764} {\bibfield  {journal}
  {\bibinfo  {journal} {Z. Phys. B Condens. Matter}\ }\textbf {\bibinfo
  {volume} {84}},\ \bibinfo {pages} {115} (\bibinfo {year} {1991})}\BibitemShut
  {NoStop}%
\bibitem [{\citenamefont {Sollich}\ and\ \citenamefont
  {Evans}(1999)}]{sollich1999glassy}%
  \BibitemOpen
  \bibfield  {author} {\bibinfo {author} {\bibfnamefont {P.}~\bibnamefont
  {Sollich}}\ and\ \bibinfo {author} {\bibfnamefont {M.~R.}\ \bibnamefont
  {Evans}},\ }\bibfield  {title} {\bibinfo {title} {Glassy time-scale
  divergence and anomalous coarsening in a kinetically constrained spin
  chain},\ }\href {https://doi.org/10.1103/PhysRevLett.83.3238} {\bibfield
  {journal} {\bibinfo  {journal} {Phys. Rev. Lett.}\ }\textbf {\bibinfo
  {volume} {83}},\ \bibinfo {pages} {3238} (\bibinfo {year}
  {1999})}\BibitemShut {NoStop}%
\bibitem [{\citenamefont {Sollich}\ and\ \citenamefont
  {Evans}(2003)}]{sollich2003glassy}%
  \BibitemOpen
  \bibfield  {author} {\bibinfo {author} {\bibfnamefont {P.}~\bibnamefont
  {Sollich}}\ and\ \bibinfo {author} {\bibfnamefont {M.~R.}\ \bibnamefont
  {Evans}},\ }\bibfield  {title} {\bibinfo {title} {Glassy dynamics in the
  asymmetrically constrained kinetic {I}sing chain},\ }\href
  {https://doi.org/10.1103/PhysRevE.68.031504} {\bibfield  {journal} {\bibinfo
  {journal} {Phys. Rev. E}\ }\textbf {\bibinfo {volume} {68}},\ \bibinfo
  {pages} {031504} (\bibinfo {year} {2003})}\BibitemShut {NoStop}%
\bibitem [{\citenamefont {Ritort}\ and\ \citenamefont
  {Sollich}(2003)}]{ritort2003glassy}%
  \BibitemOpen
  \bibfield  {author} {\bibinfo {author} {\bibfnamefont {F.}~\bibnamefont
  {Ritort}}\ and\ \bibinfo {author} {\bibfnamefont {P.}~\bibnamefont
  {Sollich}},\ }\bibfield  {title} {\bibinfo {title} {Glassy dynamics of
  kinetically constrained models},\ }\href
  {https://www.tandfonline.com/doi/abs/10.1080/0001873031000093582} {\bibfield
  {journal} {\bibinfo  {journal} {Advances in Physics}\ }\textbf {\bibinfo
  {volume} {52}},\ \bibinfo {pages} {219} (\bibinfo {year} {2003})}\BibitemShut
  {NoStop}%
\bibitem [{\citenamefont {Garrahan}\ \emph {et~al.}(2011)\citenamefont
  {Garrahan}, \citenamefont {Sollich},\ and\ \citenamefont
  {Toninelli}}]{garrahan2011kinetically}%
  \BibitemOpen
  \bibfield  {author} {\bibinfo {author} {\bibfnamefont {J.~P.}\ \bibnamefont
  {Garrahan}}, \bibinfo {author} {\bibfnamefont {P.}~\bibnamefont {Sollich}},\
  and\ \bibinfo {author} {\bibfnamefont {C.}~\bibnamefont {Toninelli}},\
  }\bibfield  {title} {\bibinfo {title} {Kinetically constrained models},\ }in\
  \href
  {http://www.oxfordscholarship.com/view/10.1093/acprof:oso/9780199691470.001.0001/acprof-9780199691470-chapter-10}
  {\emph {\bibinfo {booktitle} {Dynamical heterogeneities in glasses, colloids,
  and granular media}}},\ \bibinfo {editor} {edited by\ \bibinfo {editor}
  {\bibfnamefont {L.}~\bibnamefont {Berthier}}, \bibinfo {editor}
  {\bibfnamefont {G.}~\bibnamefont {Biroli}}, \bibinfo {editor} {\bibfnamefont
  {J.-P.}\ \bibnamefont {Bouchaud}}, \bibinfo {editor} {\bibfnamefont
  {L.}~\bibnamefont {Cipelletti}},\ and\ \bibinfo {editor} {\bibfnamefont
  {W.}~\bibnamefont {van Saarloos}}}\ (\bibinfo  {publisher} {Oxford University
  Press},\ \bibinfo {address} {Oxford},\ \bibinfo {year} {2011})\BibitemShut
  {NoStop}%
\bibitem [{\citenamefont {Roy}\ and\ \citenamefont
  {Lazarides}(2020)}]{roy2020strong}%
  \BibitemOpen
  \bibfield  {author} {\bibinfo {author} {\bibfnamefont {S.}~\bibnamefont
  {Roy}}\ and\ \bibinfo {author} {\bibfnamefont {A.}~\bibnamefont
  {Lazarides}},\ }\bibfield  {title} {\bibinfo {title} {Strong ergodicity
  breaking due to local constraints in a quantum system},\ }\href
  {https://doi.org/10.1103/PhysRevResearch.2.023159} {\bibfield  {journal}
  {\bibinfo  {journal} {Phys. Rev. Research}\ }\textbf {\bibinfo {volume}
  {2}},\ \bibinfo {pages} {023159} (\bibinfo {year} {2020})}\BibitemShut
  {NoStop}%
\bibitem [{\citenamefont {van Horssen}\ \emph {et~al.}(2015)\citenamefont {van
  Horssen}, \citenamefont {Levi},\ and\ \citenamefont
  {Garrahan}}]{horssen2015dynamics}%
  \BibitemOpen
  \bibfield  {author} {\bibinfo {author} {\bibfnamefont {M.}~\bibnamefont {van
  Horssen}}, \bibinfo {author} {\bibfnamefont {E.}~\bibnamefont {Levi}},\ and\
  \bibinfo {author} {\bibfnamefont {J.~P.}\ \bibnamefont {Garrahan}},\
  }\bibfield  {title} {\bibinfo {title} {Dynamics of many-body localization in
  a translation-invariant quantum glass model},\ }\href
  {https://doi.org/10.1103/PhysRevB.92.100305} {\bibfield  {journal} {\bibinfo
  {journal} {Phys. Rev. B}\ }\textbf {\bibinfo {volume} {92}},\ \bibinfo
  {pages} {100305} (\bibinfo {year} {2015})}\BibitemShut {NoStop}%
\bibitem [{\citenamefont {Lan}\ \emph {et~al.}(2018)\citenamefont {Lan},
  \citenamefont {van Horssen}, \citenamefont {Powell},\ and\ \citenamefont
  {Garrahan}}]{lan2018quantum}%
  \BibitemOpen
  \bibfield  {author} {\bibinfo {author} {\bibfnamefont {Z.}~\bibnamefont
  {Lan}}, \bibinfo {author} {\bibfnamefont {M.}~\bibnamefont {van Horssen}},
  \bibinfo {author} {\bibfnamefont {S.}~\bibnamefont {Powell}},\ and\ \bibinfo
  {author} {\bibfnamefont {J.~P.}\ \bibnamefont {Garrahan}},\ }\bibfield
  {title} {\bibinfo {title} {Quantum slow relaxation and metastability due to
  dynamical constraints},\ }\href
  {https://doi.org/10.1103/PhysRevLett.121.040603} {\bibfield  {journal}
  {\bibinfo  {journal} {Phys. Rev. Lett.}\ }\textbf {\bibinfo {volume} {121}},\
  \bibinfo {pages} {040603} (\bibinfo {year} {2018})}\BibitemShut {NoStop}%
\bibitem [{\citenamefont {Pancotti}\ \emph {et~al.}(2020)\citenamefont
  {Pancotti}, \citenamefont {Giudice}, \citenamefont {Cirac}, \citenamefont
  {Garrahan},\ and\ \citenamefont {Ba\~nuls}}]{pancotti2020quantum}%
  \BibitemOpen
  \bibfield  {author} {\bibinfo {author} {\bibfnamefont {N.}~\bibnamefont
  {Pancotti}}, \bibinfo {author} {\bibfnamefont {G.}~\bibnamefont {Giudice}},
  \bibinfo {author} {\bibfnamefont {J.~I.}\ \bibnamefont {Cirac}}, \bibinfo
  {author} {\bibfnamefont {J.~P.}\ \bibnamefont {Garrahan}},\ and\ \bibinfo
  {author} {\bibfnamefont {M.~C.}\ \bibnamefont {Ba\~nuls}},\ }\bibfield
  {title} {\bibinfo {title} {Quantum east model: Localization, nonthermal
  eigenstates, and slow dynamics},\ }\href
  {https://doi.org/10.1103/PhysRevX.10.021051} {\bibfield  {journal} {\bibinfo
  {journal} {Phys. Rev. X}\ }\textbf {\bibinfo {volume} {10}},\ \bibinfo
  {pages} {021051} (\bibinfo {year} {2020})}\BibitemShut {NoStop}%
\bibitem [{\citenamefont {Das}\ \emph {et~al.}(2018)\citenamefont {Das},
  \citenamefont {Chakrabarty}, \citenamefont {Dhar}, \citenamefont {Kundu},
  \citenamefont {Huse}, \citenamefont {Moessner}, \citenamefont {Ray},\ and\
  \citenamefont {Bhattacharjee}}]{das2018light}%
  \BibitemOpen
  \bibfield  {author} {\bibinfo {author} {\bibfnamefont {A.}~\bibnamefont
  {Das}}, \bibinfo {author} {\bibfnamefont {S.}~\bibnamefont {Chakrabarty}},
  \bibinfo {author} {\bibfnamefont {A.}~\bibnamefont {Dhar}}, \bibinfo {author}
  {\bibfnamefont {A.}~\bibnamefont {Kundu}}, \bibinfo {author} {\bibfnamefont
  {D.~A.}\ \bibnamefont {Huse}}, \bibinfo {author} {\bibfnamefont
  {R.}~\bibnamefont {Moessner}}, \bibinfo {author} {\bibfnamefont {S.~S.}\
  \bibnamefont {Ray}},\ and\ \bibinfo {author} {\bibfnamefont {S.}~\bibnamefont
  {Bhattacharjee}},\ }\bibfield  {title} {\bibinfo {title} {Light-cone
  spreading of perturbations and the butterfly effect in a classical spin
  chain},\ }\href {https://doi.org/10.1103/PhysRevLett.121.024101} {\bibfield
  {journal} {\bibinfo  {journal} {Phys. Rev. Lett.}\ }\textbf {\bibinfo
  {volume} {121}},\ \bibinfo {pages} {024101} (\bibinfo {year}
  {2018})}\BibitemShut {NoStop}%
\bibitem [{\citenamefont {Deger}\ \emph {et~al.}(2022)\citenamefont {Deger},
  \citenamefont {Roy},\ and\ \citenamefont {Lazarides}}]{deger2022arresting}%
  \BibitemOpen
  \bibfield  {author} {\bibinfo {author} {\bibfnamefont {A.}~\bibnamefont
  {Deger}}, \bibinfo {author} {\bibfnamefont {S.}~\bibnamefont {Roy}},\ and\
  \bibinfo {author} {\bibfnamefont {A.}~\bibnamefont {Lazarides}},\ }\bibfield
  {title} {\bibinfo {title} {Arresting classical many-body chaos by kinetic
  constraints},\ }\href {https://doi.org/10.1103/PhysRevLett.129.160601}
  {\bibfield  {journal} {\bibinfo  {journal} {Phys. Rev. Lett.}\ }\textbf
  {\bibinfo {volume} {129}},\ \bibinfo {pages} {160601} (\bibinfo {year}
  {2022})}\BibitemShut {NoStop}%
\bibitem [{\citenamefont {Kinzel}(1983)}]{kinzel1983percolating}%
  \BibitemOpen
  \bibfield  {author} {\bibinfo {author} {\bibfnamefont {W.}~\bibnamefont
  {Kinzel}},\ }\bibfield  {title} {\bibinfo {title} {Percolation structures and
  processes},\ }\href@noop {} {\bibfield  {journal} {\bibinfo  {journal} {in
  Ann. Isr. Phys. Soc., edited by G. Deutscher, R. Zallen, and J. Adler, volume
  5, Adam Hilger, Bristol}\ } (\bibinfo {year} {1983})}\BibitemShut {NoStop}%
\bibitem [{\citenamefont {Kinzel}(1985)}]{kinzel1985phase}%
  \BibitemOpen
  \bibfield  {author} {\bibinfo {author} {\bibfnamefont {W.}~\bibnamefont
  {Kinzel}},\ }\bibfield  {title} {\bibinfo {title} {Phase transitions of
  cellular automata},\ }\href
  {https://link.springer.com/content/pdf/10.1007/BF01309255.pdf} {\bibfield
  {journal} {\bibinfo  {journal} {Z. Phys. B}\ }\textbf {\bibinfo {volume}
  {58}},\ \bibinfo {pages} {229} (\bibinfo {year} {1985})}\BibitemShut
  {NoStop}%
\bibitem [{\citenamefont {Hinrichsen}(2000)}]{hinrichsen2000non}%
  \BibitemOpen
  \bibfield  {author} {\bibinfo {author} {\bibfnamefont {H.}~\bibnamefont
  {Hinrichsen}},\ }\bibfield  {title} {\bibinfo {title} {Non-equilibrium
  critical phenomena and phase transitions into absorbing states},\ }\href
  {https://doi.org/10.1080/00018730050198152} {\bibfield  {journal} {\bibinfo
  {journal} {Advances in Physics}\ }\textbf {\bibinfo {volume} {49}},\ \bibinfo
  {pages} {815} (\bibinfo {year} {2000})}\BibitemShut {NoStop}%
\bibitem [{\citenamefont {Henkel}\ \emph {et~al.}(2008)\citenamefont {Henkel},
  \citenamefont {Hinrichsen},\ and\ \citenamefont
  {L{\"u}beck}}]{henkel2008non}%
  \BibitemOpen
  \bibfield  {author} {\bibinfo {author} {\bibfnamefont {M.}~\bibnamefont
  {Henkel}}, \bibinfo {author} {\bibfnamefont {H.}~\bibnamefont {Hinrichsen}},\
  and\ \bibinfo {author} {\bibfnamefont {S.}~\bibnamefont {L{\"u}beck}},\
  }\href {https://books.google.co.in/books?id=OKbtkq4A-1EC} {\emph {\bibinfo
  {title} {Non-Equilibrium Phase Transitions: Volume 1: Absorbing Phase
  Transitions}}},\ Theoretical and Mathematical Physics\ (\bibinfo  {publisher}
  {Springer Netherlands},\ \bibinfo {year} {2008})\BibitemShut {NoStop}%
\bibitem [{\citenamefont {Broadbent}\ and\ \citenamefont
  {Hammersley}(1957)}]{broadbent1957percolation}%
  \BibitemOpen
  \bibfield  {author} {\bibinfo {author} {\bibfnamefont {S.~R.}\ \bibnamefont
  {Broadbent}}\ and\ \bibinfo {author} {\bibfnamefont {J.~M.}\ \bibnamefont
  {Hammersley}},\ }\bibfield  {title} {\bibinfo {title} {Percolation
  processes},\ }\href {https://doi.org/10.1017/S0305004100032680} {\bibfield
  {journal} {\bibinfo  {journal} {Math. Proc. Camb. Phil. Soc.}\ }\textbf
  {\bibinfo {volume} {53}},\ \bibinfo {pages} {629–641} (\bibinfo {year}
  {1957})}\BibitemShut {NoStop}%
\bibitem [{sup()}]{supp}%
  \BibitemOpen
  \href@noop {} {}\bibinfo {note} {See supplementary material at
  [URL].}\BibitemShut {Stop}%
\bibitem [{\citenamefont {{Janssen}}(1981)}]{janssen1981nonequilibrium}%
  \BibitemOpen
  \bibfield  {author} {\bibinfo {author} {\bibfnamefont {H.~K.}\ \bibnamefont
  {{Janssen}}},\ }\bibfield  {title} {\bibinfo {title} {{On the nonequilibrium
  phase transition in reaction-diffusion systems with an absorbing stationary
  state}},\ }\href {https://doi.org/10.1007/BF01319549} {\bibfield  {journal}
  {\bibinfo  {journal} {Z. Phys. B}\ }\textbf {\bibinfo {volume} {42}},\
  \bibinfo {pages} {151} (\bibinfo {year} {1981})}\BibitemShut {NoStop}%
\bibitem [{\citenamefont {{Grassberger}}(1982)}]{grassberger1982phase}%
  \BibitemOpen
  \bibfield  {author} {\bibinfo {author} {\bibfnamefont {P.}~\bibnamefont
  {{Grassberger}}},\ }\bibfield  {title} {\bibinfo {title} {{On phase
  transitions in {S}chl{\"o}gl's second model}},\ }\href
  {https://doi.org/10.1007/BF01313803} {\bibfield  {journal} {\bibinfo
  {journal} {Z. Phys. B}\ }\textbf {\bibinfo {volume} {47}},\ \bibinfo {pages}
  {365} (\bibinfo {year} {1982})}\BibitemShut {NoStop}%
\bibitem [{\citenamefont {Jensen}(1993)}]{jensen1993critical}%
  \BibitemOpen
  \bibfield  {author} {\bibinfo {author} {\bibfnamefont {I.}~\bibnamefont
  {Jensen}},\ }\bibfield  {title} {\bibinfo {title} {Critical behavior of the
  pair contact process},\ }\href {https://doi.org/10.1103/PhysRevLett.70.1465}
  {\bibfield  {journal} {\bibinfo  {journal} {Phys. Rev. Lett.}\ }\textbf
  {\bibinfo {volume} {70}},\ \bibinfo {pages} {1465} (\bibinfo {year}
  {1993})}\BibitemShut {NoStop}%
\bibitem [{\citenamefont {Jensen}(1994)}]{jensen1994critical}%
  \BibitemOpen
  \bibfield  {author} {\bibinfo {author} {\bibfnamefont {I.}~\bibnamefont
  {Jensen}},\ }\bibfield  {title} {\bibinfo {title} {Critical behaviour of a
  surface reaction model with infinitely many absorbing states},\ }\href
  {https://doi.org/10.1088/0305-4470/27/3/001} {\bibfield  {journal} {\bibinfo
  {journal} {J. Phys. A}\ }\textbf {\bibinfo {volume} {27}},\ \bibinfo {pages}
  {L61} (\bibinfo {year} {1994})}\BibitemShut {NoStop}%
\bibitem [{\citenamefont {Albano}(1995)}]{albano1995irreversible}%
  \BibitemOpen
  \bibfield  {author} {\bibinfo {author} {\bibfnamefont {E.~V.}\ \bibnamefont
  {Albano}},\ }\bibfield  {title} {\bibinfo {title} {Irreversible phase
  transitions into non-unique absorbing states in a multicomponent reaction
  system},\ }\href {https://doi.org/10.1016/0378-4371(94)00253-P} {\bibfield
  {journal} {\bibinfo  {journal} {Physica A: Statistical Mechanics and its
  Applications}\ }\textbf {\bibinfo {volume} {214}},\ \bibinfo {pages} {426}
  (\bibinfo {year} {1995})}\BibitemShut {NoStop}%
\bibitem [{\citenamefont {Mu\~noz}\ \emph {et~al.}(1996)\citenamefont
  {Mu\~noz}, \citenamefont {Grinstein}, \citenamefont {Dickman},\ and\
  \citenamefont {Livi}}]{munoz1996critical}%
  \BibitemOpen
  \bibfield  {author} {\bibinfo {author} {\bibfnamefont {M.~A.}\ \bibnamefont
  {Mu\~noz}}, \bibinfo {author} {\bibfnamefont {G.}~\bibnamefont {Grinstein}},
  \bibinfo {author} {\bibfnamefont {R.}~\bibnamefont {Dickman}},\ and\ \bibinfo
  {author} {\bibfnamefont {R.}~\bibnamefont {Livi}},\ }\bibfield  {title}
  {\bibinfo {title} {Critical behavior of systems with many absorbing states},\
  }\href {https://doi.org/10.1103/PhysRevLett.76.451} {\bibfield  {journal}
  {\bibinfo  {journal} {Phys. Rev. Lett.}\ }\textbf {\bibinfo {volume} {76}},\
  \bibinfo {pages} {451} (\bibinfo {year} {1996})}\BibitemShut {NoStop}%
\bibitem [{\citenamefont {Mu\~noz}\ \emph {et~al.}(1997)\citenamefont
  {Mu\~noz}, \citenamefont {Grinstein}, \citenamefont {Dickman},\ and\
  \citenamefont {Livi}}]{munoz1997infinite}%
  \BibitemOpen
  \bibfield  {author} {\bibinfo {author} {\bibfnamefont {M.~A.}\ \bibnamefont
  {Mu\~noz}}, \bibinfo {author} {\bibfnamefont {G.}~\bibnamefont {Grinstein}},
  \bibinfo {author} {\bibfnamefont {R.}~\bibnamefont {Dickman}},\ and\ \bibinfo
  {author} {\bibfnamefont {R.}~\bibnamefont {Livi}},\ }\bibfield  {title}
  {\bibinfo {title} {Infinite numbers of absorbing states: critical behavior},\
  }\href {https://doi.org/10.1016/S0167-2789(96)00280-1} {\bibfield  {journal}
  {\bibinfo  {journal} {Physica D: Nonlinear Phenomena}\ }\textbf {\bibinfo
  {volume} {103}},\ \bibinfo {pages} {485} (\bibinfo {year}
  {1997})}\BibitemShut {NoStop}%
\bibitem [{\citenamefont {Mu\~noz}\ \emph {et~al.}(1998)\citenamefont
  {Mu\~noz}, \citenamefont {Grinstein},\ and\ \citenamefont
  {Dickman}}]{munoz1998phase}%
  \BibitemOpen
  \bibfield  {author} {\bibinfo {author} {\bibfnamefont {M.~A.}\ \bibnamefont
  {Mu\~noz}}, \bibinfo {author} {\bibfnamefont {G.}~\bibnamefont {Grinstein}},\
  and\ \bibinfo {author} {\bibfnamefont {R.}~\bibnamefont {Dickman}},\
  }\bibfield  {title} {\bibinfo {title} {Phase structure of systems with
  infinite numbers of absorbing states},\ }\href
  {https://doi.org/10.1023/A:1023021409588} {\bibfield  {journal} {\bibinfo
  {journal} {J. Stat. Phys.}\ }\textbf {\bibinfo {volume} {91}},\ \bibinfo
  {pages} {541} (\bibinfo {year} {1998})}\BibitemShut {NoStop}%
\bibitem [{\citenamefont {Houlrik}\ \emph {et~al.}(1990)\citenamefont
  {Houlrik}, \citenamefont {Webman},\ and\ \citenamefont
  {Jensen}}]{houlrik1990meanfield}%
  \BibitemOpen
  \bibfield  {author} {\bibinfo {author} {\bibfnamefont {J.~M.}\ \bibnamefont
  {Houlrik}}, \bibinfo {author} {\bibfnamefont {I.}~\bibnamefont {Webman}},\
  and\ \bibinfo {author} {\bibfnamefont {M.~H.}\ \bibnamefont {Jensen}},\
  }\bibfield  {title} {\bibinfo {title} {Mean-field theory and critical
  behavior of coupled map lattices},\ }\href
  {https://doi.org/10.1103/PhysRevA.41.4210} {\bibfield  {journal} {\bibinfo
  {journal} {Phys. Rev. A}\ }\textbf {\bibinfo {volume} {41}},\ \bibinfo
  {pages} {4210} (\bibinfo {year} {1990})}\BibitemShut {NoStop}%
\bibitem [{\citenamefont {Liu}\ \emph {et~al.}(2021)\citenamefont {Liu},
  \citenamefont {Willsher}, \citenamefont {Bilitewski}, \citenamefont {Li},
  \citenamefont {Smith}, \citenamefont {Christensen}, \citenamefont
  {Moessner},\ and\ \citenamefont {Knolle}}]{liu2021butterfly}%
  \BibitemOpen
  \bibfield  {author} {\bibinfo {author} {\bibfnamefont {S.-W.}\ \bibnamefont
  {Liu}}, \bibinfo {author} {\bibfnamefont {J.}~\bibnamefont {Willsher}},
  \bibinfo {author} {\bibfnamefont {T.}~\bibnamefont {Bilitewski}}, \bibinfo
  {author} {\bibfnamefont {J.-J.}\ \bibnamefont {Li}}, \bibinfo {author}
  {\bibfnamefont {A.}~\bibnamefont {Smith}}, \bibinfo {author} {\bibfnamefont
  {K.}~\bibnamefont {Christensen}}, \bibinfo {author} {\bibfnamefont
  {R.}~\bibnamefont {Moessner}},\ and\ \bibinfo {author} {\bibfnamefont
  {J.}~\bibnamefont {Knolle}},\ }\bibfield  {title} {\bibinfo {title}
  {Butterfly effect and spatial structure of information spreading in a chaotic
  cellular automaton},\ }\href {https://doi.org/10.1103/PhysRevB.103.094109}
  {\bibfield  {journal} {\bibinfo  {journal} {Phys. Rev. B}\ }\textbf {\bibinfo
  {volume} {103}},\ \bibinfo {pages} {094109} (\bibinfo {year}
  {2021})}\BibitemShut {NoStop}%
\bibitem [{\citenamefont {{Willsher}}\ \emph {et~al.}(2022)\citenamefont
  {{Willsher}}, \citenamefont {{Liu}}, \citenamefont {{Moessner}},\ and\
  \citenamefont {{Knolle}}}]{willsher2022measurement}%
  \BibitemOpen
  \bibfield  {author} {\bibinfo {author} {\bibfnamefont {J.}~\bibnamefont
  {{Willsher}}}, \bibinfo {author} {\bibfnamefont {S.-W.}\ \bibnamefont
  {{Liu}}}, \bibinfo {author} {\bibfnamefont {R.}~\bibnamefont {{Moessner}}},\
  and\ \bibinfo {author} {\bibfnamefont {J.}~\bibnamefont {{Knolle}}},\
  }\href@noop {} {\bibinfo {title} {{Measurement-induced phase transition in a
  classical, chaotic many-body system}}} (\bibinfo {year} {2022}),\ \Eprint
  {https://arxiv.org/abs/2203.11303} {arXiv:2203.11303 [cond-mat.stat-mech]}
  \BibitemShut {NoStop}%
\bibitem [{\citenamefont {Kauffman}(1984)}]{kauffman1984emergent}%
  \BibitemOpen
  \bibfield  {author} {\bibinfo {author} {\bibfnamefont {S.~A.}\ \bibnamefont
  {Kauffman}},\ }\bibfield  {title} {\bibinfo {title} {Emergent properties in
  random complex automata},\ }\href
  {https://doi.org/10.1016/0167-2789(84)90257-4} {\bibfield  {journal}
  {\bibinfo  {journal} {Physica D}\ }\textbf {\bibinfo {volume} {10}},\
  \bibinfo {pages} {145} (\bibinfo {year} {1984})}\BibitemShut {NoStop}%
\bibitem [{\citenamefont {Martin}(1985)}]{martin1985lyapunov}%
  \BibitemOpen
  \bibfield  {author} {\bibinfo {author} {\bibfnamefont {O.}~\bibnamefont
  {Martin}},\ }\bibfield  {title} {\bibinfo {title} {Lyapunov exponents of
  stochastic dynamical systems},\ }\href {https://doi.org/10.1007/BF01020611}
  {\bibfield  {journal} {\bibinfo  {journal} {J. Stat. Phys.}\ }\textbf
  {\bibinfo {volume} {41}},\ \bibinfo {pages} {249} (\bibinfo {year}
  {1985})}\BibitemShut {NoStop}%
\bibitem [{\citenamefont {Derrida}\ and\ \citenamefont
  {Stauffer}(1986)}]{derrida1986phase}%
  \BibitemOpen
  \bibfield  {author} {\bibinfo {author} {\bibfnamefont {B.}~\bibnamefont
  {Derrida}}\ and\ \bibinfo {author} {\bibfnamefont {D.}~\bibnamefont
  {Stauffer}},\ }\bibfield  {title} {\bibinfo {title} {Phase transitions in
  two-dimensional {K}auffman cellular automata},\ }\href
  {https://doi.org/10.1209/0295-5075/2/10/001} {\bibfield  {journal} {\bibinfo
  {journal} {EPL (Europhysics Letters)}\ }\textbf {\bibinfo {volume} {2}},\
  \bibinfo {pages} {739} (\bibinfo {year} {1986})}\BibitemShut {NoStop}%
\bibitem [{\citenamefont {Grassberger}(1986)}]{grassberger1986spreading}%
  \BibitemOpen
  \bibfield  {author} {\bibinfo {author} {\bibfnamefont {P.}~\bibnamefont
  {Grassberger}},\ }\bibfield  {title} {\bibinfo {title} {Spreading of epidemic
  processes leading to fractal structures},\ }in\ \href@noop {} {\emph
  {\bibinfo {booktitle} {Fractals in Physics}}}\ (\bibinfo  {publisher}
  {Elsevier (Amsterdam)},\ \bibinfo {year} {1986})\ pp.\ \bibinfo {pages}
  {273--278}\BibitemShut {NoStop}%
\bibitem [{\citenamefont {Janssen}\ \emph {et~al.}(1999)\citenamefont
  {Janssen}, \citenamefont {Oerding}, \citenamefont {Van~Wijland},\ and\
  \citenamefont {Hilhorst}}]{janssen1999levy}%
  \BibitemOpen
  \bibfield  {author} {\bibinfo {author} {\bibfnamefont {H.~K.}\ \bibnamefont
  {Janssen}}, \bibinfo {author} {\bibfnamefont {K.}~\bibnamefont {Oerding}},
  \bibinfo {author} {\bibfnamefont {F.}~\bibnamefont {Van~Wijland}},\ and\
  \bibinfo {author} {\bibfnamefont {H.~J.}\ \bibnamefont {Hilhorst}},\
  }\bibfield  {title} {\bibinfo {title} {L{\'e}vy-flight spreading of epidemic
  processes leading to percolating clusters},\ }\href
  {https://doi.org/10.1007/s100510050596} {\bibfield  {journal} {\bibinfo
  {journal} {Eur. Phys. J. B}\ }\textbf {\bibinfo {volume} {7}},\ \bibinfo
  {pages} {137} (\bibinfo {year} {1999})}\BibitemShut {NoStop}%
\bibitem [{\citenamefont {Hinrichsen}\ and\ \citenamefont
  {Howard}(1999)}]{hinrichsen1999model}%
  \BibitemOpen
  \bibfield  {author} {\bibinfo {author} {\bibfnamefont {H.}~\bibnamefont
  {Hinrichsen}}\ and\ \bibinfo {author} {\bibfnamefont {M.}~\bibnamefont
  {Howard}},\ }\bibfield  {title} {\bibinfo {title} {A model for anomalous
  directed percolation},\ }\href {https://doi.org/10.1007/s100510050656}
  {\bibfield  {journal} {\bibinfo  {journal} {Eur. Phys. J. B}\ }\textbf
  {\bibinfo {volume} {7}},\ \bibinfo {pages} {635} (\bibinfo {year}
  {1999})}\BibitemShut {NoStop}%
\end{thebibliography}%

\clearpage

\onecolumngrid

\setcounter{equation}{0}
\setcounter{figure}{0}
\setcounter{page}{1}
\def\theequation{S\arabic{equation}}
\def\thefigure{S\arabic{figure}}

\begin{center}
{\textbf{{Supplementary Material: Constrained Dynamics and Directed Percolation}}}\\
Aydin Deger, Achilleas Lazarides, and Sthitadhi Roy
\end{center}

\twocolumngrid

\section{Statistical irrelevance of rare active sites}

In this section, we will discuss the additional absorbing states which have rare active sites and show that they are statistically irrelevant for the scaling near the transition. We start with arguing how these rare active sites can persist for arbitrarily large times. Here we restrict to 1+1D; generalisations to 2+1D are straightforward.

Consider a segment of  $2r+1$ sites centred at $i$, labelled by $i$, $i\pm1\cdots i\pm r$,  such that $S^z_i(t)<\cos\theta_c$ whereas $S^z_{i\pm l}(t)>\cos\theta_c$ for $l=1,\cdots,r$. This automatically means that all sites in this segment are inactive except for the sites at $i\pm 1$. This activity pattern is infinitely long-lived if the dynamics, \eqref{eq:rotation}, given by
\eq{
    \vec{S}_{i\pm 1}(t+mT) = \left[\mathsf{R_x}[\gamma_{x,i\pm1}(t)] \mathsf{R_z}[\gamma_{z,i\pm 1}(t)]\right]^m \vec{S}_{i\pm 1}(t)\,,
}
with
\eq{
    \begin{split}
    \gamma_{z,i\pm 1}(t) &= \bigg[S^z_i(t)+S^z_{i\pm 2}(t)+h\bigg]T/2\\
    \gamma_{x,i}(t) &= g T/2\\
    \end{split}\,,
}
is such that
\eq{
    S^z_{i\pm 1}(t+mT)<\cos\theta_c\quad \forall~ 1\leq m\leq n\,,
}
and 
\eq{
\left[\mathsf{R_x}[\gamma_{x,i\pm1}(t)] \mathsf{R_z}[\gamma_{z,i\pm 1}(t)]\right]^n = \mathbb{I}_3\,.
}
In other words, such an activity profile,  $\cdots 0010100\cdots$, can persist for arbitrarily long times (and hence is an absorbing state) if the dynamics of the two active spins separated by one and surrounded by inactive spins are such that they form a $n$-cycle and the two spins stay inside the constraining sector throughout the cycle. We next provide numerical evidence for this mechanism.

\begin{figure}[!t]
    \centering
    \includegraphics[width=0.8\linewidth]{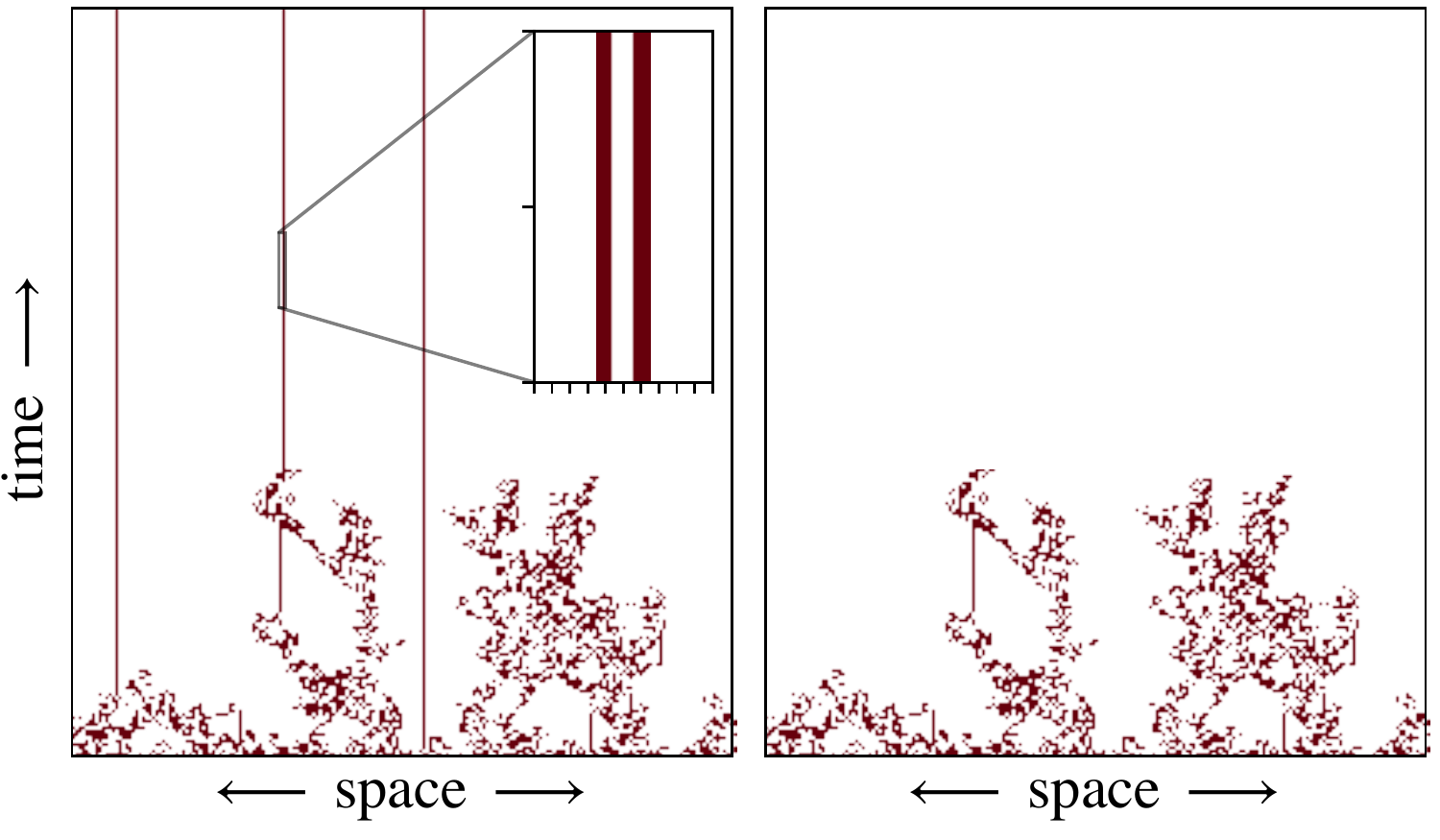}
    \caption{Left: The activity map for a randomly chosen initial condition the absorbing state for which has rare active sites. The inset shows a zoom which shows that the active sites appear in isolated pairs with a single inactive site separating them. Right: The same map as the left panel except the persistent active sites have been identified and rendered inactive. Results for L=1024 and $\theta_c/\pi = 0.53$.}
    \label{fig:tower}
\end{figure}

In Fig.~\ref{fig:tower}(left) we show the activity map for a randomly chosen initial state the dynamics for which exhibits such rare active sites in the absorbing state.
The inset shows that the active sites indeed appear in pairs separated by one inactive site in the middle and surrounded also by inactive sites.
\begin{figure}
    \centering
    \includegraphics[width=0.7\linewidth]{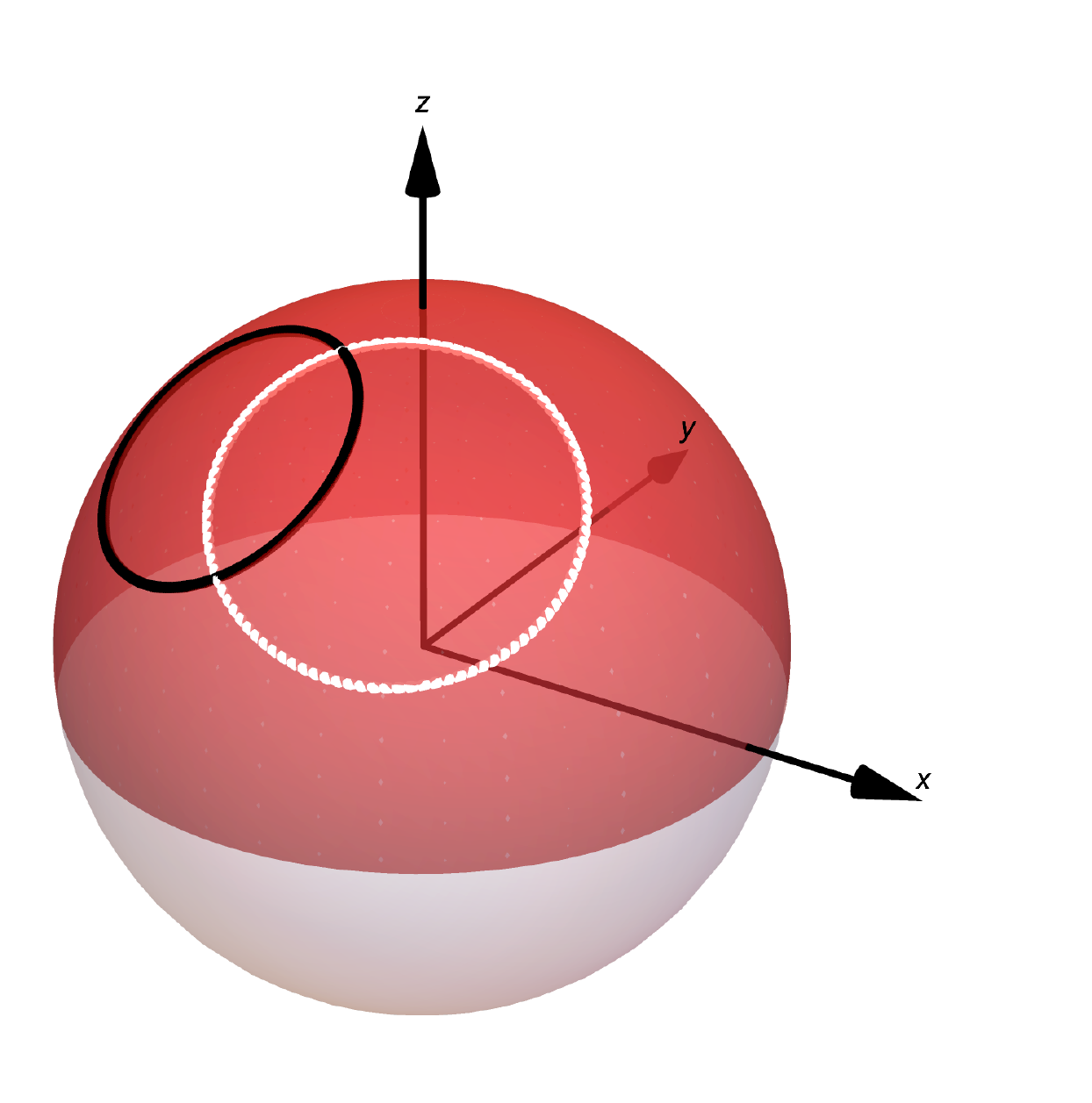}
    \caption{The trajectories of two active spins (black and white) on the unit sphere which correspond to the ones shown in the inset of Fig.~\ref{fig:tower}(left), form cycles while staying completely within the constrained sector denoted by the red shade.}
    \label{fig:cycle}
\end{figure}
In Fig.~\ref{fig:cycle}, we plot the trajectories of the two active spins show in the inset of Fig.~\ref{fig:tower}(left) on the unit sphere. The trajectories indeed form closed curves which reside entirely in the constrained sector of the sphere indicated by the red shade. These two numerical results show that the mechanism discussed above is indeed the one that gives rise to additional absorbing states with active sites.

\begin{figure}[!t]
    \centering
    \includegraphics[width=\linewidth]{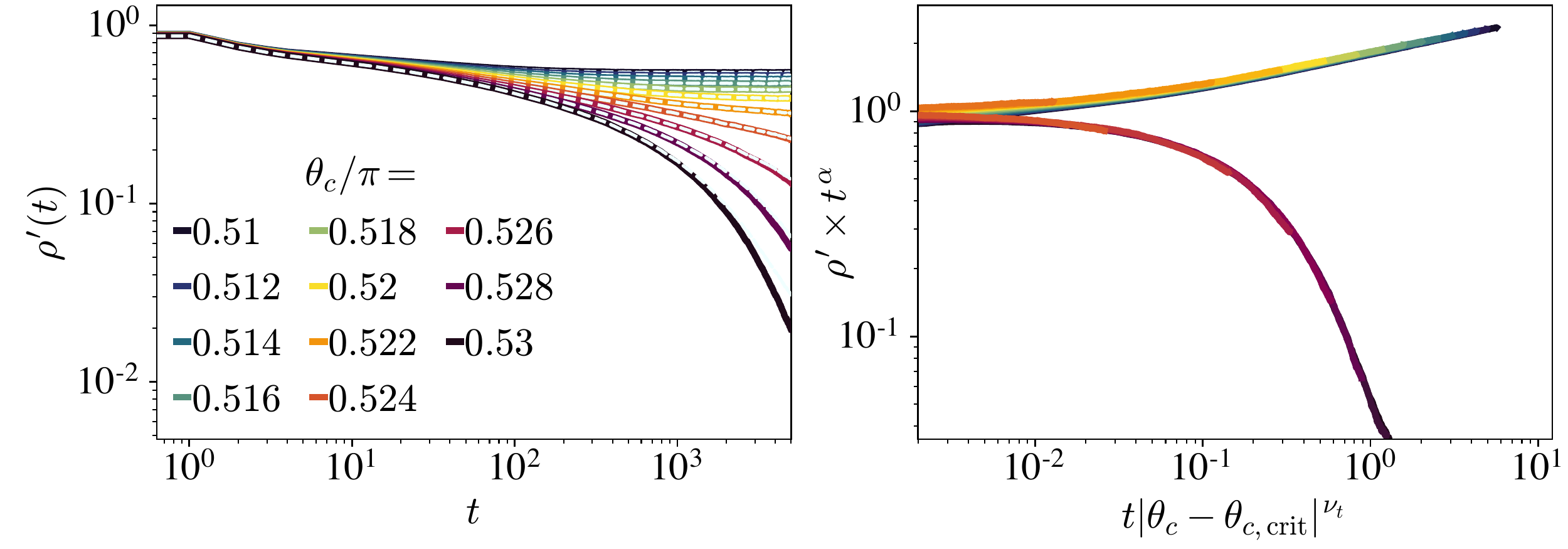}
    \caption{Results for modified density of active sites, $\rho^\prime(t)$ where rare persistent active sites in the absorbing states are rendered inactive by hand. The white dashed lines show the corresponding data for $\rho(t)$ which are almost indistinguishable. The scaling collapse with the DP exponents, as in Fig.~\ref{fig:1dheis}, is again excellent indicating that the rare active sites are irrelevant for the reults of the scaling analysis. Results for $L=2048$.}
    \label{fig:rhoprime}
\end{figure}

Let us now discuss the implications (or absence thereof) of these absorbing states for the scaling analyses. In the main text, the results for the average density of active sites and its scaling analyses treated these additional absorbing states on the same footing as the absorbing state with $\sigma_i=0~\forall~ i$; to compute the average $\rho(t)$ an unbiased average was performed over random initial conditions whose absorbing states may or may not contain these rare absorbing states.

What we find is that the results remain virtually unchanged if these persistent active sites are treated as inactive sites. The rationale behind this is that in the frozen phase, these active sites cannot grow into larger percolating clusters and the activity pattern of $\cdots 01010 \cdots$ persists -- they are effectively inactive sites as far as cluster growth is considered. To make this quantitative, we identify these active sites for each initial condition by searching for activity patterns of the form $\cdots 01010\cdots$ which persist for the final one-tenth of the total simulation time. We then consider the quintet and locate the time at which their activity pattern changed for the last time and set them to be inactive for all times after that. The right panel in Fig.~\ref{fig:tower} illustrates the result of this process for the activity map shown in the left panel; the activity map stays the same except the persistent active sites are rendered inactive. Following this process, we again compute the average density of active sites, and denote it by $\rho^\prime(t)$ [the primed notation distinguishes it from the original $\rho(t)$]. We show the results for $\rho^\prime(t)$ and its scaling analysis in Fig.~\ref{fig:rhoprime}. The results are almost identical, quantitatively, to those of $\rho(t)$ and consequently, show excellent scaling collapse with the DP exponents as in Fig.~\ref{fig:1dheis}. This shows that the additional absorbing states with rare active sites are irrelevant for the scaling analyses near the transition.

\section{Stochastic Constrained Dynamics}

\begin{figure}[!b]
    \centering
    \includegraphics[width=\linewidth]{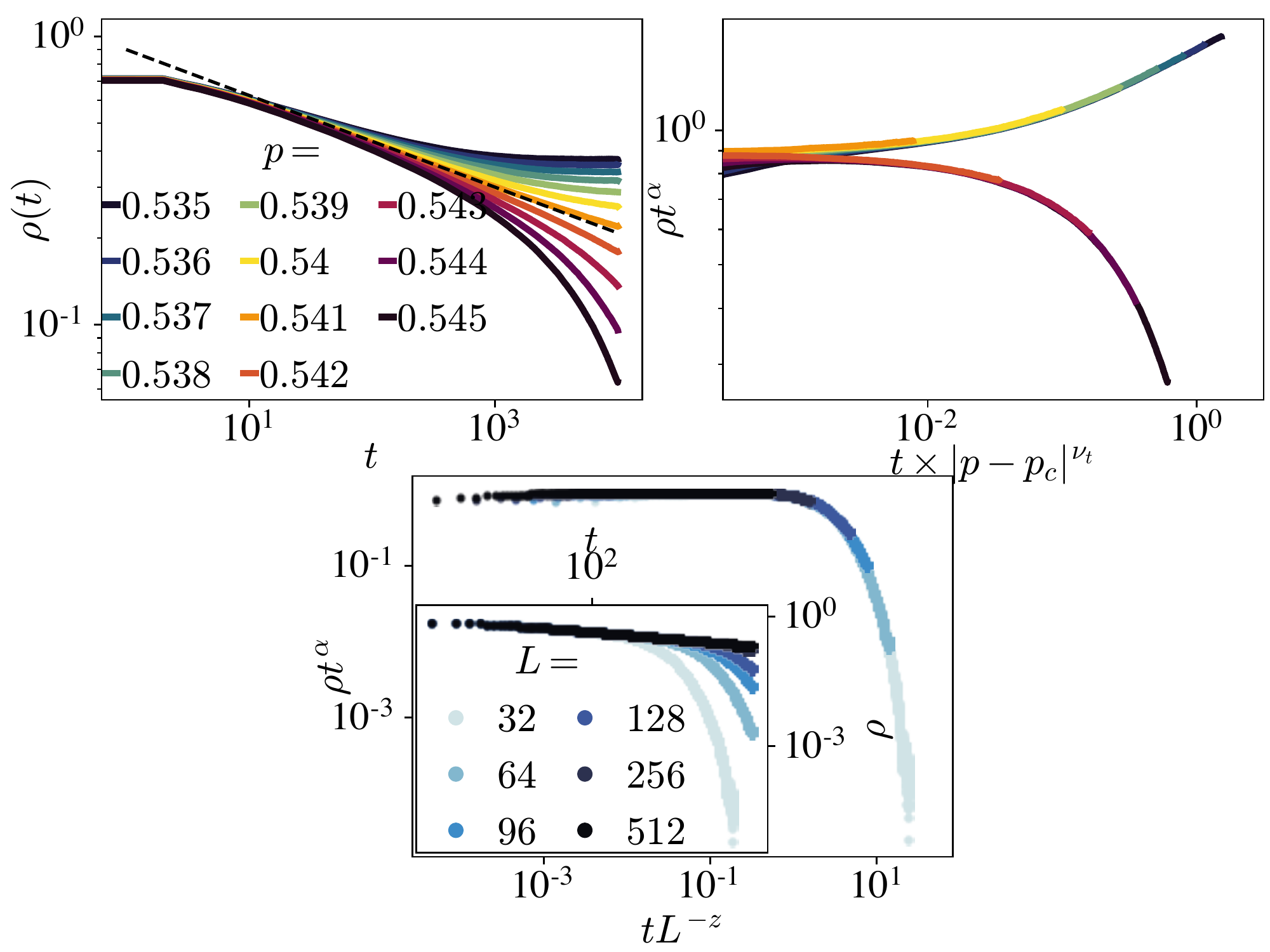}
    \caption{Results for the scaling analyses for the stochastic constrained dynamics defined in Eq.~\eqref{eq:stoch}. Top left: The density of active sites, $\rho$, as a function of $t$ for different values of $p$. At the critical point, $\rho$ decays as a power law, $t^{-\alpha}$ with  $\alpha = 0.159$ consistent with DP in 1+1D (black dashed line). Top right: Plotting $\rho t^\alpha$ as a function of $t|p-p_c|^{\nu_t}$ with $p_c=0.5413$ and the DP exponents $\alpha=0.159$ and $\nu_t = 1.734$ shows a perfect collapse. Bottom: Finite-size scaling at the critical point by plotting $\rho t^\alpha$ against $tL^{-z}$ with the DP universality exponent, $z=1.581$, again shows excellent collapse.}

    \label{fig:stoch}
\end{figure}

In this section, we discuss the 1+1D stochastic model discussed in the main text as a coarse-grained version of the spin system. We find that this stochastic model also has a dynamical phase transition which lies in the DP universality class.

The dynamics is defined in terms of two Boolean variables, $\tau_i$ and $\sigma_i$, on site $i$. $\tau_i=1/0$ is a proxy for the spin at site $i$ being outside/inside the constraining cone and $\sigma_i \equiv \mathsf{OR}[\tau_{i-1},\tau_{i+1}]$ is a proxy for the activity of the site. The stochastic dynamics are given by
\eq{
    \tau_{i}(t+1) = \sigma_{i}(t)\eta_{i,t} + [1-\sigma_i(t)]\tau_i(t)\,,
    \label{eq:stoch}
}
where $\eta_{i,t}$'s are independent random numbers,
\eq{
  \eta_{i,t} = \begin{cases}
  0;~~~& \mathrm{probability}~p \\
  1;~~~& \mathrm{probability}~1-p
  \end{cases}\,.
}
The dynamical rules in Eq.~\eqref{eq:stoch} imply that if a site is active at time $t$, then at time $t+1$ it goes into the constraining cone with probability $p$ and out of the cone with probability $1-p$.
 The probability $p$ then in some sense plays the same role as the  angle $\theta_c$. A larger $\theta_c$ would mean that it is more likely that under dynamics the spin enters the cone implying a larger $p$. Therefore, for this model, $p$ plays the role of the tuning parameter for the transition.

 As in the main text, we start from an all active initial condition, $\tau_i(0)=1$ for all $i$ which also implies $\sigma_i(0)=1$. We run the dynamical rules over several realisations and compute the average activity $\rho(t) = N^{-1}\sum_{i}\braket{\sigma_i(t)}$. We perform the same scaling analyses as in the main text; the results are shown in Fig.~\ref{fig:stoch}. The excellent scaling collapse of the data with the DP exponents (see Tab.~\ref{tab:exponents}) provides clear evidence that the transition in this stochastic model lies in the DP universality class.

\end{document}